\documentclass[preprint2]{aastex}

\usepackage{natbib}
\usepackage{graphicx}
\usepackage{multirow}
\usepackage{txfonts}

\begin{document}

\shorttitle{Accretion Flow Dynamics of MAXI~J1836-194 with TCAF Solution}
\shortauthors{A. Jana, D. Debnath, S. K. Chakrabarti et al.}

\title{ACCRETION FLOW DYNAMICS OF MAXI J1836-194 DURING ITS 2011 OUTBURST FROM TCAF SOLUTION}
\author{Arghajit Jana\altaffilmark{1}, Dipak Debnath\altaffilmark{1}, Sandip K. Chakrabarti\altaffilmark{2,1}, Santanu Mondal\altaffilmark{1}, Aslam Ali Molla\altaffilmark{1}}
\altaffiltext{1}{Indian Center for Space Physics, 43 Chalantika, Garia St. Rd., Kolkata, 700084, India.}
\altaffiltext{2}{S. N. Bose National Centre for Basic Sciences, Salt Lake, Kolkata, 700098, India.}

\email{argha@csp.res.in, dipak@csp.res.in; chakraba@bose.res.in; santanu@csp.res.in, aslam@csp.res.in}


\begin{abstract}

The Galactic transient X-ray binary MAXI~J1836-194 was discovered on 29th August 2011.
Here we make a detailed study of the spectral and timing properties of its 2011 outburst 
using archival data of RXTE Proportional Counter Array instrument. The evolution of accretion 
flow dynamics of the source during the outburst through spectral analysis with Chakrabarti-Titarchuk's 
two-component advective flow (TCAF) solution as a local table model in XSPEC. We also 
fitted spectra with combined disk blackbody and power-law models and compared it with the TCAF model fitted results. 
The source is found to be in hard and hard-intermediate spectral states only during entire phase of this outburst.
No soft or soft-intermediate spectral states are observed. This could be due to the fact that this object belongs 
to a special class of sources (e.g., MAXI~J1659-152, Swift~J1753.5-0127, etc.) that have very short 
orbital periods and that companion is profusely mass-losing or the disk is immersed inside an excretion disk. 
In these cases, flows in the accretion disk is primarily dominated by low viscous sub-Keplerian 
flow and the Keplerian rate is not high enough to initiate softer states. Low-frequency quasi-periodic 
oscillations (QPOs) are observed sporadically although as in normal outbursts of transient black holes, 
monotonic evolutions of QPO frequency during both rising and declining phases are observed. 
From the TCAF fits, we find mass of the black hole in the range of $7.5-11~M_\odot$ and from time differences 
between peaks of the Keplerian and sub-Keplerian accretion rates we obtain viscous timescale for this particular 
outburst $\sim 10$~days.

\end{abstract}



\section{Introduction}

Compact objects such as neutron stars, black holes, etc. are characterized by electromagnetic radiations emitted 
from the accretion disks, forms due to the accreting matter supplied by their companions. Some of these objects are 
transient binaries in nature and are very interesting to study using X-rays as they undergo rapid evolutions in 
their spectral and timing properties. 
Several works are present in the literature (see, e.g., Tomsick et al., 2000; McClintock \& Remillard, 2006; 
Debnath et al., 2008, 2013; Nandi et al., 2012; Rao, 2013) to explain variation of timing and spectral 
properties of these objects during their X-ray outbursts. 
It also has been reported by several authors that these objects exhibit different spectral states 
(for e.g., hard, hard-intermediate, soft-intermediate and soft) during their outbursts (see, Debnath et al., 2013 
and references therein). Low frequency quasi-periodic oscillations (QPOs) are often observed in the 
power-density spectra (PDS) of some of these spectral states (see Remillard \& McClintock 2006 for a review). 
In the literature, many authors (Belloni et al., 2005; Debnath et al., 2013 and references therein) have reported 
that model extracted physical parameters undergo a hysteresis loop during their spectral evolution through 
an entire epoch of the outburst of these transient black hole candidates (BHCs).

It is well-known that the emitted spectrum of radiation coming from BHCs is multicolor in nature and contains 
both nonthermal and thermal components. The thermal component is the multicolor blackbody radiation from the 
standard Keplerian disk (Shakura \& Sunyaev, 1973; Novikov \& Thorne, 1973) and the other is the power-law component 
that originates from a ``Compton" cloud (Sunyaev \& Titarchuk 1980, 1985), a repository of hot electrons whose
thermal energy is transferred to low-energy photons from the standard disk by repeated Compton scatterings to 
produce high-energy X-rays. In the two-component advective flow (TCAF) solution of Chakrabarti \& Titarchuk (1995, hereafter CT95; 
see also, Chakrabarti, 1997), the so-called {\it `Compton cloud'} or {\it `hot corona'} is actually the CENtrifugal 
pressure supported BOundary Layer (CENBOL) that automatically forms behind the centrifugal barrier due to 
the pileup of the low viscosity (lesser than critical viscosity) and low angular momentum optically thin matter 
known as sub-Keplerian (halo) accretion component. An axisymmetric shock (Chakrabarti, 1990, 1996; Ryu et al., 1997) 
defines the outer boundary of the CENBOL. Okuda et al., (2007) showed that this shock is stable even for 
non-axisymmetric perturbations. In TCAF another component of the accretion flow is the viscous, optically thick 
and geometrically thin Keplerian (disk) component which is submerged inside the sub-Keplerian (halo) component. 
CENBOL being much hotter in general, the Keplerian disk is naturally truncated at the shock location close to 
the black hole. This Keplerian flow settles down to a standard SS73 disk when cooling is efficient 
(see, Giri et al. 2015, and references therein). 

Recently, after the inclusion of two-component advective flow (TCAF) solution in HEASARC's spectral analysis software 
package XSPEC (Arnaud, 1996) as a local additive table model (Debnath, Chakrabarti \& Mondal, 2014, hereafter DCM14; 
Mondal, Debnath \& Chakrabarti, 2014a, hereafter MDC14; Debnath, Mondal \& Chakrabarti, 2015a, hereafter DMC15; 
Debnath, Molla, Chakrabarti \& Mondal, 2015b, hereafter DMCM15), which requires only five physical parameters 
(including mass) to fit a spectrum from a BHC, an accurate picture of the accretion flow dynamics around several 
transient BHCs (e.g., H~1743-322, GX~339-4, MAXI~J1659-152) during their X-ray outbursts was obtained. From the 
TCAF model fitted spectrum one cannot only directly obtain information about instantaneous accretion rates of
the two components but also obtain crucial information on shock parameters (instantaneous shock location $X_s$ and shock
strength $\beta=1/R$, where $R$ is the shock compression ratio) which allow us to estimate frequencies of the dominating QPOs 
(if observed in PDS; see, DCM14). Various spectral states are observed during an outburst phase of a transient BHC, 
can be characterized by interesting variations of the {\it accretion rate ratio} (ARR) and QPOs (shape, frequency, $Q$ value, and rms\%). 
This motivated us to study the accretion flow dynamics of newly discovered Galactic BHC MAXI~J1836-194 
during its very first outburst in 2011 with TCAF, particularly because of its very low orbital timescale.

The transient BHC MAXI~J1836-194 has a short orbital period of $< 4.9$~hrs and a low disk inclination angle 
($4-15^\circ$; Russell et al., 2014). The source located at R.A. $= 18^h35^m43.43^s$, Dec $= -19^\circ 19'12.1''$ was 
first observed simultaneously by SWIFT/BAT and MAXI/GSC on 2011 August 29 (Negoro et al., 2011). Miller-Jones et al. (2011), 
based on X-ray spectral analysis and observation of a radio jet when a transition from hard to soft state occurs, 
suggested the source to be a potential BHC. Reis et al. (2012) suggested the source to be a highly rotating BHC with 
a spin parameter of $a=0.88\pm0.03$. VLT optical spectral study of the binary system by Russell et al. (2014) suggests 
that the mass, and the distance of the source to be $4-12~M_\odot$, $4-10$~kpc respectively. They also computed the 
mass  and the radius of the companion donor main sequence star to be $< 0.65~M_\odot$, and $<0.59~R_\odot$ respectively. 
On the contrary, Cenko et al. (2011) concluded the companion to be a high-massive Be star on the basis of their 
optical spectroscopic study and system as a high-mass X-ray binary. 

MAXI~J1836-194 showed its first X-ray flaring activity on 29th August 2011 (Modified Julian Day, i.e., MJD=55802), which 
continued for the next $\sim 2$~months. During this outburst period source was studied extensively in multi-energy bands 
such as various X-ray (Ferrigno et al., 2012; Reis et al., 2012; Radhika et al., 2014), and radio observatories 
(Yang et al., 2012; Russell et al. 2013). An evolving radio jet was also observed by Russell et al. (2013). 
We study timing and spectral properties of the BHC during the entire phase of the outburst using RXTE 
Proportional Counter Array (PCA) archival data. 
In the present work we give the TCAF fitted spectral analysis results to study accretion flow dynamics around the BHC 
during its 2011 outburst and discuss how the ARR values and nature of QPOs (if present) vary with spectral states.
We also compare our TCAF model fitted spectral results with that of the combined disk blackbody (DBB) 
and power-law (PL) model fitted results.

The {\it paper} is organized in the following way. In \S 2, we briefly discuss observation and data 
analysis procedures using HEASARC's HeaSoft software package. In \S 3, we present spectral analysis results 
using the TCAF solution based {\it fits} file as a local additive table model in XSPEC. 
We also fitted spectra with the DBB model plus the PL model and compared analysis results with those of the TCAF fitted 
spectral results to demonstrate the strength of TCAF fits, to understand different spectral states and their 
correlation with temporal properties. We show that on some days there are 
clear indications of unresolved double peaked iron lines. Finally in \S 4, a brief discussion on the results 
and concluding remarks are presented.

\section{Observation and Data Analysis}

RXTE/PCA monitored the source two days after its discovery on a daily basis (Strohmayer \& Smith 2011) 
except for a crucial $\sim 5$~days gap during MJD=$55813-18$ in the rising phase of the outburst. 
Here we study achival data of $35$ observational IDs starting from the first PCA observed day; 2011 August 31 
(MJD=55804) to 2011 November 24 (MJD = 55889) using XSPEC version 12.8. 
To analyze the data we follow standard data analysis technique of the RXTE/PCA instrument 
as presented in Debnath et al. (2013, 2015a). 

For timing analysis we use the PCA {\it Science Binned} mode (FS3f*.gz) data with a maximum timing
resolution of $125\mu s$ to generate light curves for well-calibrated Proportional Counter Unit 2 
(PCU2; including all six layers) in $2-25$~keV (0-59 channels) and $2-15$~keV (0-35 channels). 
To generate the PDS ``powspec" routine of the XRONOS package is used to compute rms fractional variability 
on $2-15$~keV light curves of $0.01$ sec time bin. 
To find centroid frequencies of QPOs we fit PDS with Lorentzian profiles and use the ``fit err" command
to get error limits.

For spectral analysis we use {\it Standard2} mode Science Data (FS4a*.gz) of the PCA instrument. 
The $2.5-25$ keV background subtracted Proportional Counter Unit 2 (PCU2) spectra are fitted with both 
the TCAF-based model {\it fits} file and combined DBB and PL model components in XSPEC. The individual 
DBB and PL model components fluxes are obtained by using the convolution model `cflux' technique. 
To achieve the best spectral fits, a Gaussian line of peak energy around $6.5$~keV (iron-line emission) 
is used except for the five observations (see, Appendix Table I), where we fitted the data with the {\it LAOR} model (Laor 1991). 
We fit the data with line energies at $\sim 7.1$ ~keV and with emissivity indices around $\sim 3.5$ 
(see, Appendix Table II). Hydrogen column density (N$_{H}$) was kept frozen at 2.0$\times$10$^{21}$~atoms~cm$^{-2}$ 
(Kennea et al., 2011) for absorption model {\it wabs}. 
For entire outburst, we also use a fixed $1.0$\% systematic instrumental error for the spectral analysis.
The XSPEC command `err' is used to find 90\% confidence `+ve' and `-ve' error values for the model fitted 
parameters after achieving the best fit based on reduced chi-square value ($\chi^2_{red} \sim 1$).
In Appendix Table I, average values of these two $\pm$ errors are mentioned in the superscripts of the 
parameter values. In Appendix Table II, we present TCAF with LAOR model fitted LAOR parameters 
for five spectra, where LAOR model is found to be more useful to deal with Fe line instead of single 
Gaussian $\sim 6.5$~keV.

To fit spectra using the TCAF-based model additive table {\it fits} file, one needs to supply five model 
input parameters such as, $i)$ black hole mass ($M_{BH}$) in solar mass ($M_\odot$) unit, $ii)$ Keplerian accretion rate 
($\dot{m_d}$ in Eddington rate $\dot{M}_{Edd}$), $ii)$ sub-Keplerian accretion rate ($\dot{m_h}$ in $\dot{M}_{Edd}$), 
$iv)$ location of the shock ($X_s$ in Schwarzschild radius $r_g$=$2GM_{BH}/c^2$), 
$v)$ compression ratio (R=$\rho_+ / \rho_-$, where $\rho_+$ and $\rho_-$ are post- and pre-shock densities respectively) 
of the shock. The model normalization value is a fraction of $\frac{r_g^2}{4\pi D^2} sin(i)$, where `$D$' is the source distance 
(in 10~kpc unit) and `$i$' is the disk inclination angle. In order to fit a black hole spectrum with TCAF model, 
we create an additive table model {\it fits} file ({\it TCAF.fits}) using theoretical model spectra that are 
generated by varying five input parameters in the modified CT95 code. 
Basically the governing equations are taken from CT95 as far as hydrodynamics and radiative transfer is concerned. 
However, the shock strength and the equation to get shock height were generalized to capture features with wider range 
of accretion rates (see, DCM14 and DMC15). In this Paper for the spectral analysis of MAXI~J1836-194 with the TCAF we keep 
mass a free parameter and find that it comes out to be in the range of $7.5-11~M_\odot$. 

\section{Results}

Our recent study shows that the accretion flow dynamics around a transient BHC can be well understood by the 
analysis of the spectral and temporal behaviors of the BHC during its outburst under TCAF model paradigm. Here 
spectral analysis results based on the TCAF model are presented. We also compare the DBB model plus the PL model 
fitted results with that of the TCAF model fitted spectral analysis results. It is to be noted that combined 
PL and DBB model fitted spectral analysis, only provides gross properties of the accretion disk such as fluxes 
from different thermal and nonthermal components, where as TCAF model goes one step further. The TCAF model 
extracts detailed physical flow parameters, such as two types of accretion rates and the Compton cloud properties. 
Furthermore, transitions between different spectral states are more conspicuous when described in terms of the fitted 
parameters as we shall also see in the case of the present object. Thus, studying accretion dynamics around BHCs 
with TCAF solution-based {\it fits} file provides us a certain definite advantages.

To study the timing and spectral properties of MAXI~J1836-194 during its current outburst, RXTE/PCA data for $35$ 
observations spread over the entire period of the 2011 outburst are used. We first studied X-ray count rate 
variation over the entire outburst phase with PCU2 light curves from all observations in the $2-25$~keV energy band 
(see Fig. 1a). From the nature of this outburst profile we may define the source into the class belonging to
a {\it fast-rise and slow-decay} type (FRSD) rather than a complete-outbursting {\it slow-rise and slow-decay} (SRSD) 
type such as GRO~J1655-40, GX~339-4, etc. (Debnath et al., 2010). QPOs are found for only $14$ observations out of 
total of $35$. For spectral study we initially fitted spectra with the combined PL and DBB model components. 
The spectral model fitted parameters such as DBB temperature ($T_{in}$ in keV), PL photon index ($\Gamma$) and 
fluxes from both model components are obtained. All the spectra are then refitted with the current version (v0.3)
of our TCAF model {\it fits} file in XSPEC to extract the accretion flow parameters such as Keplerian disk rate ($\dot{m_d}$), 
sub-Keplerian halo rate ($\dot{m_h}$), shock location ($X_s$) and the compression ratio ($R$) of the shock. 

\subsection{Spectral Analysis with the TCAF Solution and with Combined DBB and PL models}

Spectral fit with the conventional DBB plus PL model, which provides us with a rough estimate of the flux 
contributions coming from both nonthermal (from PL) and thermal (from DBB) processes around a BH. 
It gives us an rough idea about the evolution of spectral states by monitoring variations of 
$T_{in}$, $\Gamma$ factors, and flux contribution from both model components. 
However, variations of the TCAF fitted/derived parameters (such as two types of accretion rates, 
$\dot{m_d}$ and $\dot{m_h}$; shock parameters, $X_s$ and $R$; and ARR) provide us with a clear picture of 
the geometry and the accretion flow dynamics around the BH during the outburst phase. In Appendix Table I, 
all these spectral fitted/derived parameters are mentioned in a tabular form with estimated errors in superscript. 
Most importantly, since the entire flow dynamics is determined by these few parameters only, we do not need 
to change the normalization constant from day to day. The uncertainity in normalization merely reflects 
the uncertainity in mass measurements. 

Figures 1-4 show variations of QPO frequencies, X-ray intensities, and spectral (with DBB plus PL model components 
and the TCAF model) fitted and derived parameters. In Figure 1a, variation of the background subtracted PCU2 count 
rate in $2-25$~keV energy band with time (day in MJD) is shown. In Figure 1c, variation of TCAF model fitted total 
accretion rates ($\dot{m_d}$ plus $\dot{m_h}$) in the energy band of $2.5-25$~keV are shown. To compare with 
PCU2 rates and with total accretion rates in Figure 1b, total flux variation of the DBB plus PL model 
fitted spectra of $2.5-25$~keV are shown. Here we observe that the variation of the TCAF fitted total flow 
rate (Fig. 1c) is different from the count rate or total flux variations in Figs. 1(a-b) especially in early 
stages of the outburst. In Fig. 1d, variation of {\it accretion rate ratio} i.e., ARR (defined as the ratio 
between sub-Keplerian halo rate with Keplerian disk rate, i.e., $\dot{m_h}$/$\dot{m_d}$) is shown. 
In Fig. 1e, we show QPO frequencies (only dominating primary) with day (in MJD). From the variations of these 
physical flow parameters, QPO frequencies, etc., only two spectral classes such as {\it hard} (HS), and 
{\it hard-intermediate} (HIMS) are observed during the entire phase of the current outburst of MAXI~J1836-194.
Note that softer states such as soft-intermediate (SIMS), and soft state (SS) are absent in contrast to what 
most of the other outburst sources exhibited (see, e.g., Nandi et al., 2012; Debnath et al., 2013). 
A similar result of no SS was observed by Debnath et al. (2015b) for another MAXI source, namely MAXI~J1659-152 
(during its 2010 outburst). The sequence in the present source appears to be: HS (rising) $\rightarrow$ HIMS (rising) 
$\rightarrow$ HIMS (declining) $\rightarrow$ HS (declining). 

In Figs. 2(a-b), variation of DBB temperature ($T_{in}$ in keV) and power-law photon index ($\Gamma$) with day (MJD) 
are shown. In Figs. 2(c-d), TCAF model fitted shock location ($X_s$ in $r_g$ unit) and compression ratio ($R$) are 
plotted with day (in MJD). In Figs. 3(a-b), variations of DBB flux from DBB plus PL model fits and TCAF model 
fitted Keplerian disk rate with day (MJD) are compared. Similarly, in Figs. 3(c-d), we compare variations of PL flux 
with sub-Keplerian halo rate from these respective models. Clearly, one can observe `some' similarities in each 
pair of these compared quantities, but not totally since the PL flux is a function of the disk rate as well.

Figure 4 shows plots of (a) hardness intensity diagram and of (b) ARR and $2-15$~keV PCU2 count rate is drawn to find 
a correlation between temporal and spectral properties of the BHC and judge their behavior vis-a-vis classifications. 
There is clearly a hysteresis behavior. This will be discussed in detail in \S 3.3. In Fig. 5, TCAF model fitted 
combined spectra (top panel) and residuals for four spectra of observation Ids: 96371-03-03-01 (MJD=55808.34, 
black online), 96438-01-01-04 (MJD=55818.85, red online), 96438-01-02-04 (MJD=55823.82, blue online), and 
96438-01-06-02 (MJD=55850.87, orange online), selected from different regions i.e., states of the outburst 
are shown. Here (i), and (iv) plots are for hard states of and plots (ii), and (iii) are for hard-intermediate 
states of rising and declining phases respectively. We observe that in (i) and (iii) the fits are better with 
an additional {\it LAOR} model described iron line component which appears to be unresolved double lines 
(Fabian et al. 1989; Stella 1990). In (ii) and (iv) we required only a single Gaussian to model the single iron lines.

To check correlation among spectral fitted parameters, in Fig. 6(a-d), we make comparative plots of 
$i)$ total (DBB plus PL) flux vs. total flow rate ($\dot{m_d}$ plus $\dot{m_h}$), $ii)$ ARR vs. QPO frequency, 
$iii)$ $\dot{m_h}$ vs. $\dot{m_d}$, and $iv)$ PL photon index $\Gamma$ vs. $\dot{m_h}$. In Fig. 6a, correlation is 
expected. In Fig. 6b, no correlation is expected in any obvious sense and the QPOs are related to 
resonance effects between cooling timescale (a function of $\dot{m_h}$ and $\dot{m_d}$) and infall timescale
(a function of $X_s$ and $R$). In Fig. 6c, weak coupling  between the two rates ($\dot{m_h}$ and $\dot{m_d}$) 
is due to the fact that at least a fraction of the disk rate is the bi-product of the halo rate mediated by viscosity.
In Fig. 6d, a weak coupling of rising of $\Gamma$ with $\dot{m_h}$ was expected even from the
basic work of CT95 when $\dot{m_d}$ is almost constant and small. 
We made statistical analysis using two correlation methods such as Spearman Rank (SR) and Pearson Linear (LP), 
and found that total fluxes are strongly correlated with total accretion rates (SR coefficient $SR_{e}$=0.964, 
LP coefficient $LP_{e}$=0.976), no or weak anti-correlation between ARR and QPO frequency ($SR_{e}$=-0.008, 
$LP_{e}$=-0.363), and a weak correlation between two types ($\dot{m_h}$, and $\dot{m_d}$) of accretion rates 
($SR_{e}$=-0.336, $LP_{e}$=-0.418), and PL photon index $\Gamma$ with halo ($\dot{m_h}$) rate ($SR_{e}$=0.271, $LP_{e}$=0.221). 
We also find that $\Gamma$ and ARR are weakly anti-correlated with each other 
($SR_{e}$=-0.278, $LP_{e}$=-0.404) as expected from the argument given above.

\subsection{Evolution of Spectral and Temporal Properties during the Outburst}

The spectral and temporal properties of this object during its first outburst are discussed by several authors on 
the basis of X-ray or radio variability, QPO observations, spectral results based on inbuilt XSPEC model fits, such as, 
thermal DBB and nonthermal PL components (Ferrigno et al. 2012; Radhika et al. 2014; Russell et al. 2014).
However, spectral fits on a daily basis with TCAF provides us with a variation of physical parameters and enlighten 
us with physical reason behind different spectral states observed during the outburst. It also allows us to 
find a pattern to correlate with spectral transitions and their sequences. Recently, interesting correlations 
in 2010 outburst of H~1743-322 (MDC14), in 2010-11 outburst of GX~339-4 (DMC15) and in 2010 outburst of 
MAXI~J1659-152 (DMCM15) were found. It was observed that two-component accretion rates ($\dot{m_d}$, $\dot{m_h}$),  
ARRs, shock locations, and compression ratios in conjunction with observed QPOs provided a better characterization of 
the classification of spectral states. The nature of different spectral states observed during the 2011 outburst of 
MAXI~J1836-194 will be discussed below in the sequence of their appearance.

\noindent{\it (i) Hard State (Rising phase):}
RXTE started observing the source two days after its discovery. The source is observed in the hard state 
for the first $\sim 7$ days of the observation (from  MJD = 55804.52 to 55810.29). Low values of the observed power-law 
photon indices (between $\sim 1.65-1.75$) allow us to conclude that the object is in the hard state. PCA count rate, 
total flow/accretion rates increase with time (day) and reach their individual maximum observable values on the transition 
(HS to HIMS) day (MJD=55810.29), when a local maximum of ARR and sub-Keplerian halo rate ($\dot{m_h}$) are also observed 
(see Figures 1 \& 3). QPOs are observed only for two observations during this phase and the frequencies 
increased as in normal outburst, indicating that the underlying shocks whose oscillations cause QPOs may be moving inward
but the resonance condition (that cooling and infall timescales should be close to each other) is not satisfied on some days.
One reason for this may be that the ARR is almost constant (see, Mondal et al., 2015, and Chakrabarti et al., 2015 for more details). 
Another reason may be that there are excess sub-Keplerian halo matter coming from the companion which is not 
a part of the accretion process that surrounds the system as a whole. It is also possible that the 
companion is a Be star and the accretion disk is immersed inside the excretion disk, a fact borne out by 
generally very low values of spectral indices in hard state spectra signifying the presence of a `super-hard' 
state. The flux is also maximum in this state, which is unusual as the flux is generally highest in the 
soft/sot-intermediate states that are missing for this outburst of the source 
(see, Debnath et al., 2008, 2013; Nandi et al., 2012). 

We classify HIMS into two parts, HIMS (rising) and HIMS (declining), since there is a sharp change of physical 
(both temporal and spectral) properties on a day that is termed as the transition day. On this day QPO 
frequency was found to be the highest.

\noindent{\it (ii) Hard-Intermediate State (Rising phase):}
This state lasts for $\sim 10$ days. ARR decreases monotonically due to rise in $\dot{m_d}$ and fall in 
$\dot{m_h}$. At the same time, PL $\Gamma$ increases monotonically and DBB $T_{in}$ decreases. This may be 
because of more supply of matter from thermally cooler Keplerian disk component as viscosity rises. 
On the transition day (MJD=55820.41). The ARR is found to have its lowest value with a maximum in $\dot{m_d}$ and 
a minimum in $\dot{m_h}$. During this phase, QPO is observed on only three days. QPO frequency continues to 
increase. This behavior may also be due to the same reason as that of the rising HS stated above.

\noindent{\it (iii) Hard-Intermediate State (Declining phase):}
For the next $\sim 11$ days starting from the transition day when the QPO frequency has the maximum value, 
source was in this spectral state. During this phase, QPOs are observed sporadically though the frequency monotonically 
decreased from $5.175$ to $2.023$~Hz within the first $\sim 9$ days of this state. ARR is found to increase 
rapidly with a rise in halo rate and decrease in disk rate (see Figure 1 and 3). This causes a rapid decrease 
in $\Gamma$ also indicates that spectrum start to become harder from the first day of this state. On September 27 
(MJD=55831.85), transition from declining hard-intermediate to hard spectral state was seen as on this 
particular day and a local maximum in the ARR is also observed. A similar nature was found on the rising 
hard to hard-intermediate transition day. Precisely this behavior was seen in our earlier study on 
other transient BHCs with the TCAF fits as well (see, MDC14, DMC15, DMCM15).

\noindent{\it (iv) Hard State (Declining phase):}
The source is found to be in this spectral state untill the end of our observation (MJD=55889.16) starting from 
the HIMS-HS transition day. In this phase of the outburst, supply from both components of matter is cut off 
as a result of that the ARR values decreasing steadily with time (day). Shock recededs back as day progresses 
with a slow rise in compression ratios. QPOs are observed sporadically during only three observations with 
the frequency decreasing monotonically. On other days the resonance condition was not fulfilled and QPOs 
were not observed. Note the unusually low value of the spectral index. This could be because the entire disk 
system is immersed in an excretion disk or winds of the companion. This is also a reason why QPOs are not 
observed regularly everyday as in the same states of other outburst sources.

\subsection{ARR-Intensity Diagram (ARRID): A Correlation Between Timing and Spectral Properties}

In Fig. 4(a-b), we showed evolution of various quantities during the entire outburst to impress that the 
parameters of the declining phase do not retrace themselves during the rising phase. In the so-called hardness-intensity (HID)
diagram, data are plotted mainly using photon counts from light curves in selected constant energy ranges throughout 
the outburst and one tries to find correlation. Sometimes the curve looks like a `q' and the diagram is called 
q-diagram (Maccarone \& Coppi, 2003; Fender et al., 2004; Belloni et al., 2005). In Fig. 4a, the HID diagram is drawn. 
Days when state transitions occur are not obvious. Since state transitions are necessarily due to accretion rate 
variations (CT95), it is instructive to make a plot with physical parameters where transitions of states are obvious 
from the diagram itself. In Fig. 4b, we plot the variation of the $2-25$~keV PCU2 count rate as a function of 
ARR ($\dot{m_h}$/$\dot{m_d}$). We define it as the {\it accretion rate ratio intensity diagram} (ARRID). 
Here B, C, and D represent the days when state transitions took place and points A, and E mark start and 
end of the observation. It clearly shows a hysteresis effect and different branches are associated with 
different spectral states. A-B (black online), D-E (violet online) represent HS in rising and declining phases 
respectively. The points B and D mark transition days between HS and HIMS, when ARR acquires local 
maximum value, irrespective of the count rate. The point C marks the transition day between two HIMS, when 
a sharp change in physical (spectral and temporal) properties, such as minimum ARR and maximum QPO frequency, 
etc. are observed, irrespective of PCU2 count rate. The approximate horizontal lines, B-C (red online) and 
C-D (green online) represent HIMS in rising and declining phases respectively. 

\section{Discussions and Concluding Remarks}

We study the evolution of the spectral and temporal properties of a Galactic transient BHC MAXI~J1836-194 
during its first (2011) X-ray outburst using RXTE PCU2 data. For spectral evolution study, we use two models 
(TCAF and DBB plus PL models) applied to a total of $35$ observations. A combined PL and DBB model fit 
provides us with a rough estimate of thermal (DBB) and nonthermal (PL) flux contributions as well as an idea 
about the disk temperatures ($T_{in}$) and PL photon indices ($\Gamma$). The TCAF model (v0.3) fit provides us 
with physical accretion flow parameters such as the Keplerian disk ($\dot{m_d}$), sub-Keplerian halo ($\dot{m_h}$) 
rates, the shock location ($X_s$), and the shock compression ratio ($R$; see, Figures 1-3). In Appendix Table I, 
a detailed spectral analysis results with observed QPO frequencies are presented. The variation of these model 
fitted parameters with the ARR (a ratio between $\dot{m_h}$ to $\dot{m_d}$), and properties of QPOs (if observed) 
provide us with a better understanding of the accretion flow dynamics around the BHC during its outburst.  

So far we have successfully studied the accretion flow dynamics of three Galactic BHCs (H~1743-322, GX~339-4, and MAXI~J1659-194) 
during their X-ray outbursts from our spectral study with the TCAF model (see MDC14, DMC15, DMCM15). A strong correlation 
between spectral and timing properties are found for these sources giving rise to transitions between different spectral states. 
Generally four (HS, HIMS, SIMS, and SS) spectral states are observed during an outburst of transient 
BHC in the sequence of HS$\rightarrow$HIMS$\rightarrow$SIMS$\rightarrow$SS$\rightarrow$SIMS$\rightarrow$HIMS$\rightarrow$HS.
According to CT95, Ebisawa et al., (1996), an outburst of transient BHCs is triggered due to a sudden rise 
in viscosity at the outer edge of the disk and the declining phase begins when the viscosity is turned off 
at the outer edge (CT95; Ebisawa et al., 1996). If viscosity does not rise above a critical value (Chakrabarti, 1990), 
the Keplerian component does not form and as a result we may miss soft states (SS) altogether during an 
outburst (DMCM15). The resulting event could be termed as a `failed' outburst. 

The QPOs (generally type `C') are found to evolve monotonically in rising/declining hard and hard-intermediate spectral states 
(Chakrabarti et al., 2005, 2008; Debnath et al., 2010, 2013; Nandi et al., 2012). We also observed a local maximum of 
ARR on the transition day of HS $\rightleftharpoons$ HIMS and of evolving QPO frequency on HIMS$\rightleftharpoons$SIMS 
transition day (MDC14, DMC15, DMCM15). QPOs are observed sporadically on and off during SIMS and are absent in SS 
(Debnath et al., 2008, 2013; Nandi et al., 2012). During the rising hard state, the ARR is found to 
increase monotonically with a rise in the halo rate and reaches its maximum on the HS-HIMS transition day and the shock is 
found to move in gradually weakening it strength. In rising HIMS the shock further moves in rapidly due to shrinking 
of the CENBOL size. This may be due to the rapid rise in the thermally cooler Keplerian disk rate, which reaches maximum 
on HIMS-SIMS transition day. Also, on the transition day, shock compression ratio ($R$) becomes approximately unity. 
During this process the ARR rapidly reduces to a lower value on the transition day. During rising SIMS, ARR are observed 
at a lower value with very little variation and generally type B or A QPOs are observed sporadically. The origin of these 
QPOs is assumed to be different and may be due to non-satisfaction of Rankine-Hugoniot condition to form a stable shock 
location (Ryu et al., 1997; Chakrabarti et al., 2015). A strong jet may be found during this phase of the outburst that 
may be due to the cooling down of the base of the CENBOL with a sudden rise in the Keplerian matter contributions. 
The cooling reduces sound speed, making the upper part of the jet become supersonic quite abruptly. In the soft state 
the ARR reduces further with a clear dominance of the Keplerian component. In the declining SIMS the contribution of 
the Keplerian component starts to decrease and sporadic appearances of QPOs are seen. The ARR varies in a similar way as 
of rising SIMS. On the declining SIMS-HIMS transition day the maximum value of the QPO frequency could be found. 
In the process, during the declining HIMS the ARR rises rapidly and reaches its maximum value on the HIMS-HS transition day. 
QPO frequencies are observed to decrease continuously in a rapid manner with an outward movement of the shock and a rise in 
compression ratios. In the declining hard state, both ARR and QPO frequency decrease with time with a rapid outward 
movement of shock. At the same time a decrease in both components of accretion rates could be observed.  

The present analysis of the outburst data reveals a great deal of surprises. Depending on the nature of the variation 
of the TCAF model fitted or derived parameters ($\dot{m_d}$, $\dot{m_h}$, ARR, $X_s$, R, etc.), and nature of the QPOs 
(if present), only two spectral classes, namely, {\it hard} (HS), and {\it hard-intermediate} (HIMS), 
are observed during the entire phase of the current outburst of MAXI~J1836-194. These spectral states are observed in the 
sequence of: HS (rising) $\rightarrow$ HIMS (rising) $\rightarrow$ HIMS (declining) $\rightarrow$ HS (declining). 
The so-called `q'-diagram does not show a `q' shape at all (see, Fig. 4a). In the ARRID diagram (Fig. 4b), the evolution 
of the $2-25$~keV PCU2 count rate with the ARR, which hints at a hysteresis behavior, strongly justifies our spectral classification. 
Different branches of the plot appear to be related to different spectral states. We clearly see that the ARR has a local maximum 
on transition days of HS $\rightleftharpoons$ HIMS. This plot provides us with a better understanding of the strong correlation 
between spectral and temporal properties. In MDC14, a similar type of correlation between different spectral states was 
also found for Galactic BHC H~1743-322 during its 2010 outburst. 

A peculiarity of this outburst of MAXI~J1836-194 is that QPOs are not seen every day, although there is 
a general trend similar to other outbursts: increasing of frequency during the rising phase and decreasing of frequency 
in the declining phase of the outburst. Also, we have not found any signature of the soft and soft-intermediate spectral 
states. The source started and ended from/to hard spectral states via two hard-intermediate spectral states during the entire 
phase of the outburst. Furthermore, the spectral indices in few observation of hard states are very low 
(i.e., spectra are super-hard). The unusual behavior of the source during this outburst may be due to a shorter 
orbital period of the binary system which makes the disk very small in size and it is possibly immersed in the wind 
or the excretion disk of the companion; this would be the case if the companion is a Be Star, for instance 
(Cenko et al., 2011). During most of this outburst the source was covered with low angular 
momentum matter which is not necessarily a part of the accretion disk. The non-observation of QPOs on a regular 
basis is also explained since resonance condition for QPOs would be difficult to fulfill in the presence of such 
extraneous matter (see, Molteni et al., 1996; Mondal et al., 2015; Chakrabarti et al., 2015) which are not included 
in the TCAF solution.

For a detailed analysis of the TCAF parameters, one may be able to estimate the viscous timescale using time differences 
between peaks of the Keplerian disk and sub-Keplerian halo component rates during rising or declining phases of an outburst 
of a transient BHC. MAXI~J1836-194 shows double humps for both of the accretion rates during the entire phase of the current outburst. 
During the rising phase the halo rate attains its maximum value on MJD=55810, whereas the disk rate attains its peak after 
$\sim 10$~days when the halo reaches its first minimum. Again, during the declining phase the halo rate starts to increase and 
attains its second peak on MJD=55826. At the same time the disk rate attains its second peak on MJD=55836, which is 
coincidentally exactly after $\sim 10$~day interval. The observed time gaps between the peaks of $\dot{m_h}$ and $\dot{m_d}$ 
in both the rising and declining phases suggests that the viscous timescale of MAXI~J1836-194 during 2011 outburst could 
be around of $\sim 10$~days.

Recently, Shaposhnikov \& Titarchuk (2009) have found that the scaling method of PL photon Index ($\Gamma$) vs. 
QPO frequency diagram is a powerful tool to predict mass of an unknown BHC. For that, one needs sufficient number 
of observations, especially at the bending and horizontal saturation branches of the diagram. This method 
is not useful to predict the mass of the current BHC MAXI~J1836-192, since there are only two observations of QPOs 
($4.35$ and $5.17$~Hz) at the saturation level, and no QPOs in the region where the slope turns into the saturation 
level for higher frequencies and indices. So we obtain the mass of the unknown BH from our TCAF fits.

To further emphasize the advantage of the TCAF fits, we wish to emphasize that the model normalization 
is the normalization of the whole spectrum that it arises out of the difference between the calculated spectrum at 
the disk frame and the observer's frame which includes the idiosyncrasies of the instruments. Once the spectrum is 
calculated using certain inner edges and other parameters there is no way that a normalization constant should 
contain the disk parameters anymore. Strangely, our precision fitting with the TCAF model does not require one to vary 
the normalization constant as in other models. This is because the TCAF solution normalization depends on the 
source distance ($D$) and the disk inclination angle ($i$), for a given mass of the black hole, all intrinsic 
properties of the disk (such as the inner edge of the disk) which are used inside normalization in other models are
all used up to obtain the spectrum itself. Indeed, we get excellent fits keeping the normalization constant within 
a narrow range as it should be for the current source MAXI~J1836-194 and the source MAXI~J1659-152 
during its 2010 outburst (Molla et al. 2016). On the other hand, the normalization required for DBB varies from day 
to day. These are given in Appendix Table I. This nonconstancy has been variously explained by the change of the inner edge 
of the disk, which we find intriguing since the truncated disk information should have already been used to fit the
data in the first place. Our model normalization is found to vary in a narrow
range of $0.25-0.35$, (DBB model normalization shows wild variation in the range of $0.34-291$)
except for five observations around the transition day between two HIMS when a prominent jet 
is observed (see, Russell et al., 2013). The deviation of the normalization in these days reflects the non-inclusion 
of contribution from the jet in the current version (v0.3) of the TCAF model {\it fits} file. During the spectral fitting, 
as we kept mass of the BH as a free parameter, the $M_{BH}$ is found in the range of $7.5-11~M_\odot$. The detailed method 
of the prediction of the BH mass from the TCAF fitting are discussed in Molla et al. (2016). 

In the future, we will make detailed spectral and temporal study of other similar short orbital period BHCs (for e.g., 
XTE~J1118+480 of orbital period $\sim 4.1$~hrs, Gonz\'{a}lez-Hern\'{a}ndez et al., 2013; Swift~J1753.5-0127 of 
orbital period $\sim 3.2$~hrs, Zurita et al., 2007) during their X-ray outbursts to check if flow dynamics of 
these sources also follow a similar trend. 

\section*{Acknowledgments}

AJ and DD acknowledge support from ISRO sponsored RESPOND project fund (ISRO/RES/2/388/2014-15). 
DD also acknowledges support from DST sponsored Fast-track Young Scientist project fund (SR/FTP/PS-188/2012).
AAM and SM acknowledge supports of MoES sponsored junior research fellowship and post-doctoral fellowship respectively.

{}

\clearpage


\begin{figure}
\vskip -0.5cm
\centerline{
\includegraphics[scale=0.6,angle=0,width=8.0truecm]{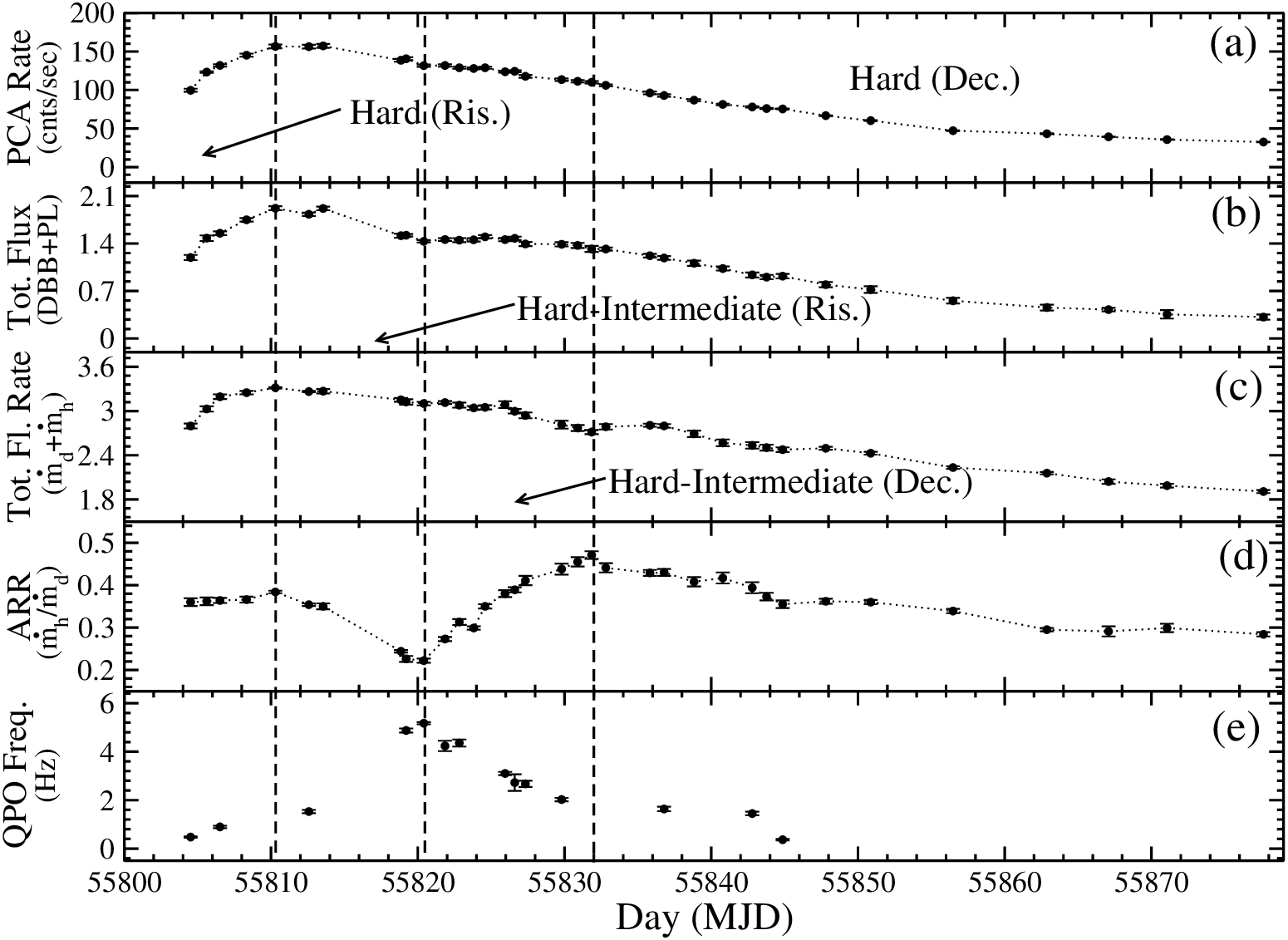}}
\caption{Variation of (a) $2-25$~keV PCA count rates (cnts/sec), (b) combined disk black body (DBB) and power-law
(PL) model fitted total spectral flux in $2.5-25$~keV range (in units of $10^{-9}~ergs~cm^{-2}~s^{-1}$),
(c) TCAF model fitted total flow (accretion) rate (in $\dot{M}$$_{Edd}$; sum of Keplerian disk, $\dot{m_d}$ and
sub-Keplerian halo $\dot{m_h}$ rates) in the $2.5-25$~keV energy band, and (d) accretion rate ratio (ARR; ratio
between halo and disk rates) with day (MJD) for the 2011 outburst of MAXI J1836-194 are shown.
In the bottom panel (e), observed primary dominating QPO frequencies (in Hz) with day (MJD) are shown.
The vertical dashed lines indicate transitions between different spectral states.}
\label{fig1}
\end{figure}

\begin{figure}
\vskip -0.5cm
\centerline{
\includegraphics[scale=0.6,angle=0,width=8.0truecm]{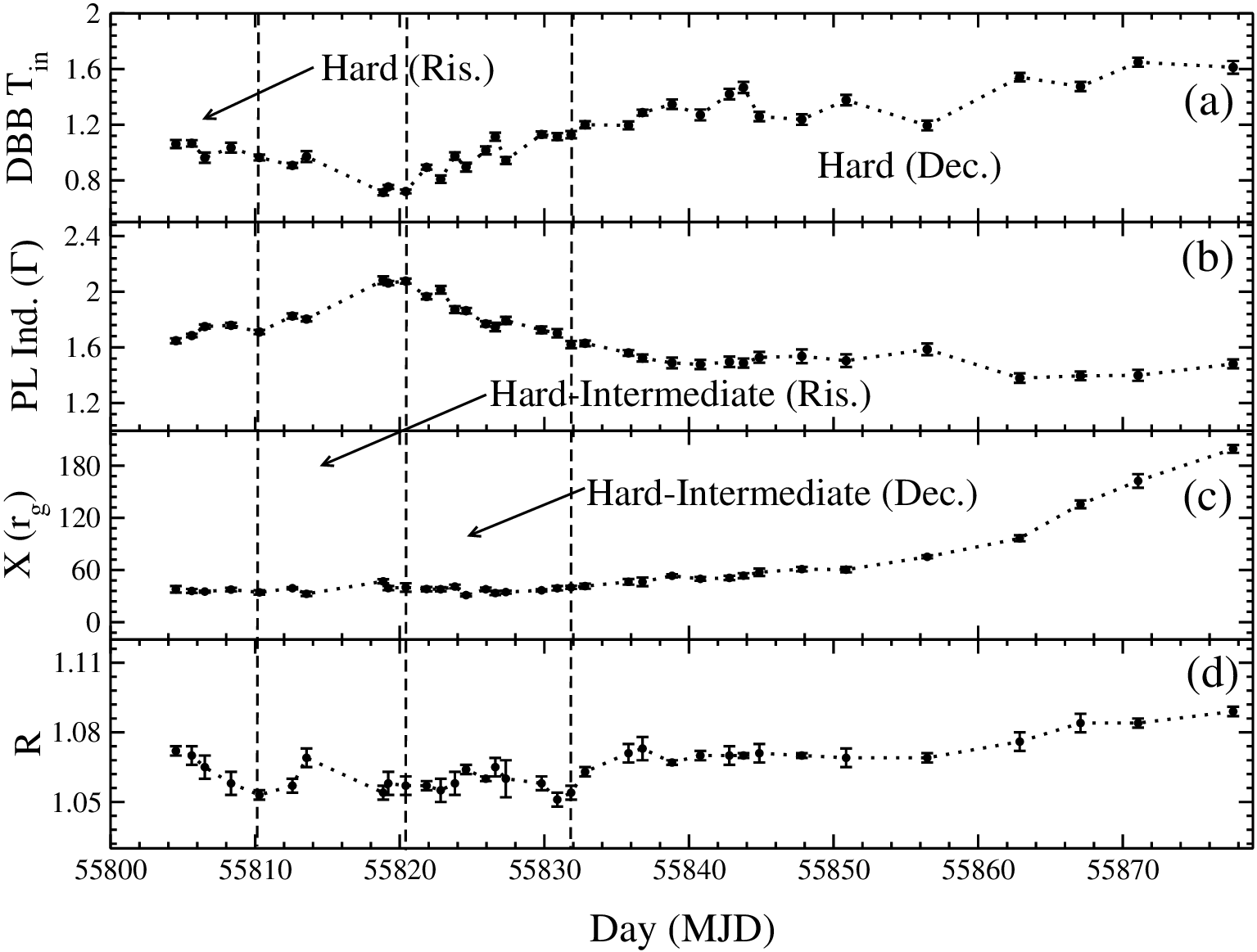}}
\label{fig2}
\caption{Variation of combined DBB and PL model fitted (a) disk temperature $T_{in}$ (in keV), and (b) 
PL photon index ($\Gamma$) with day (MJD) are shown in top two panels. Variations of TCAF model fitted/derived 
(c) shock location ($X_s$ in $r_g$) and (d) compression ratio ($R$), with day (MJD) are shown.
}
\end{figure}

\begin{figure}
\vskip -0.5cm
\centerline{
\includegraphics[scale=0.6,angle=0,width=8.0truecm]{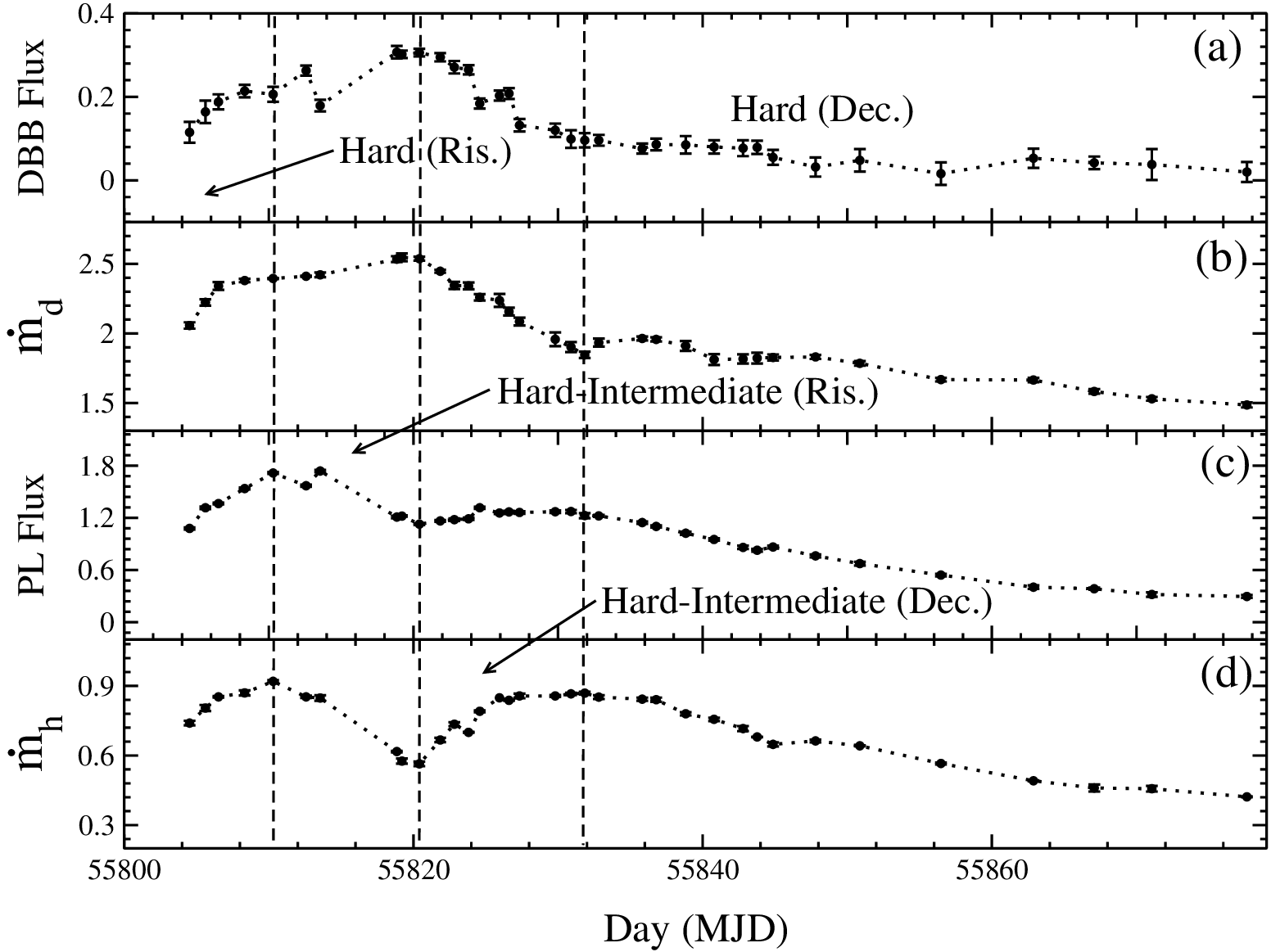}}
\label{fig3}
\caption{In top panel (a), the variation of combined disk black body (DBB) and power-law (PL) model fitted DBB
spectral flux and in panel (c), the variation of PL spectral flux (both in units of $10^{-9}~ergs~cm^{-2}~s^{-1}$)
in $2.5-25$~keV energy range are shown. In panel (b), the variation of TCAF model fitted Keplerian disk rate $\dot{m_d}$
and in bottom panel (d), the variation of sub-Keplerian halo rate $\dot{m_h}$ (both in $\dot{M}$$_{Edd}$) in the
same energy band are shown.
}
\end{figure}

\begin{figure}
\vskip -0.5cm
\centerline{
\includegraphics[scale=0.6,angle=0,width=4.2truecm]{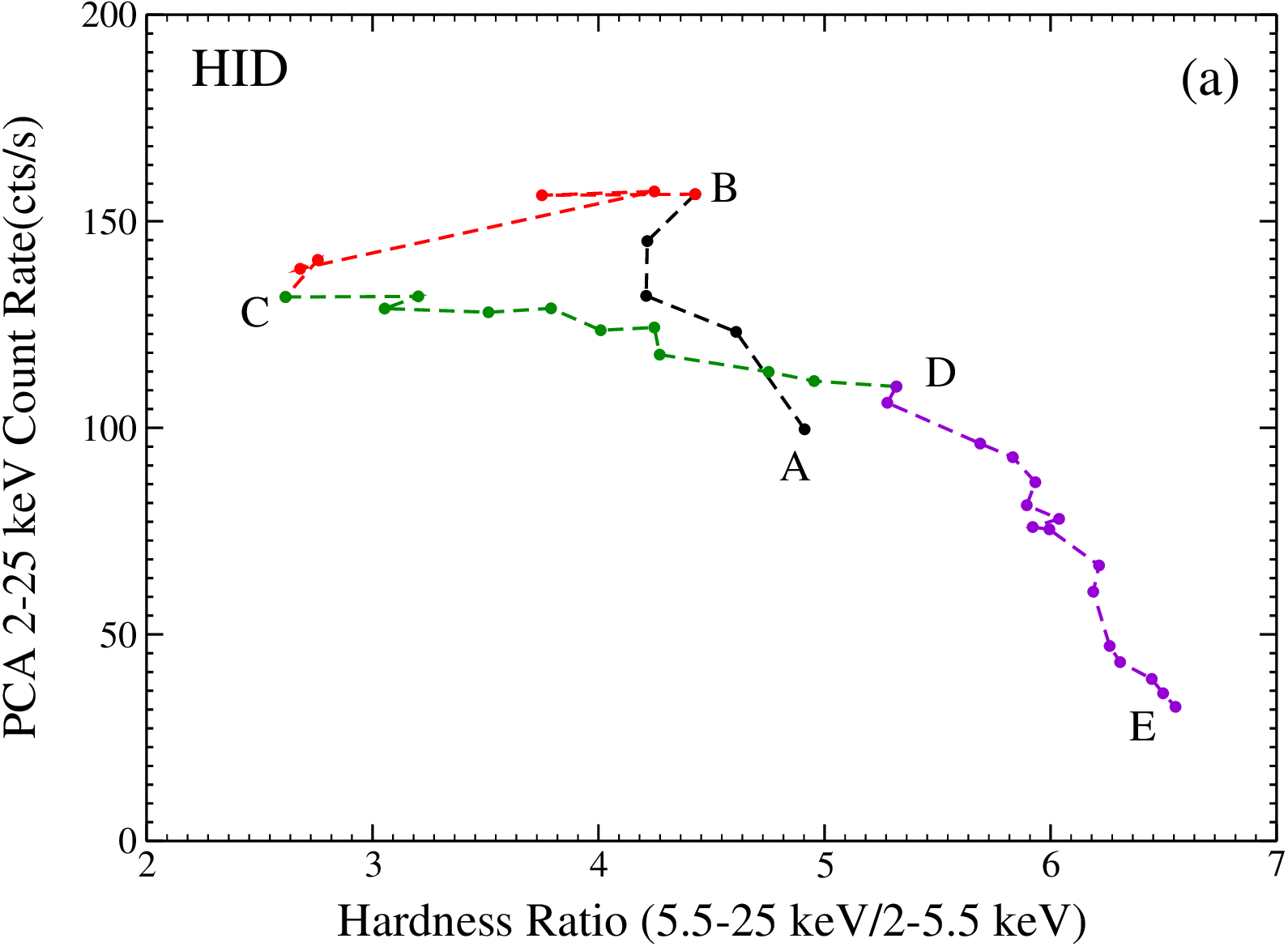}\hskip 0.2cm
\includegraphics[scale=0.6,angle=0,width=4.2truecm]{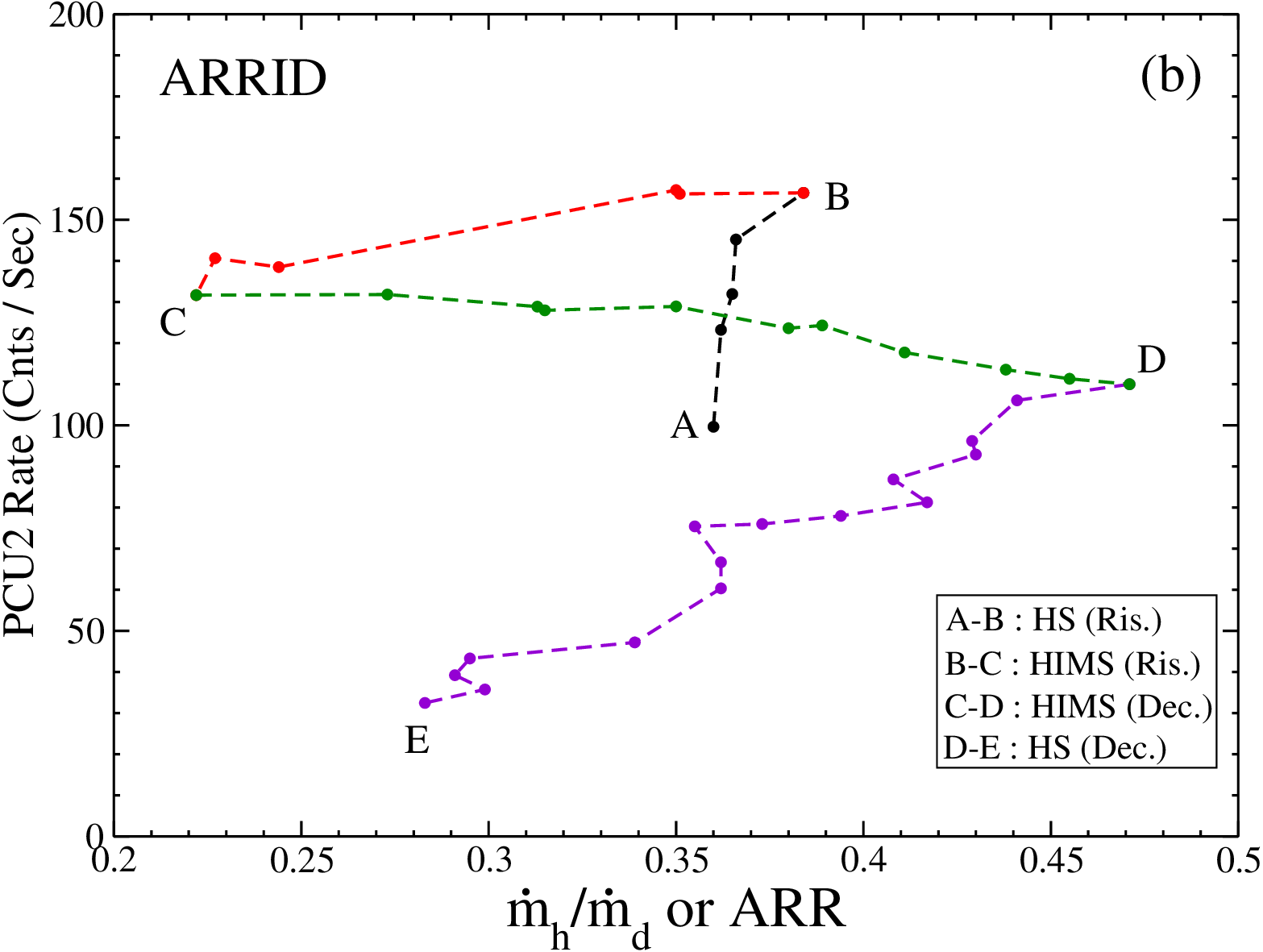}
}
\caption{(a) Hardness (ratio between $5.5-25$~keV to $2-5.5$~keV PCU2 count rates)-Intensity diagram 
for the entire 2011 outburst of MAXI~J1836-194. Note that the 'q' shape (for e.g., Belloni et al., 2005) is not 
formed in this case. (b) Evolution of $2-15$~keV PCA count rate as a function of ARR ($\dot{m_h}$/$\dot{m_d}$),
showing the hysteresis effect. Transitions from HS to HIMS (rising) and the reverse
(declining) take place when ARR is local maximum. 
}
\label{fig4}
\end{figure}

\begin{figure}
\vskip -0.5cm
\centerline{
\includegraphics[scale=0.6,angle=270,width=8.2truecm]{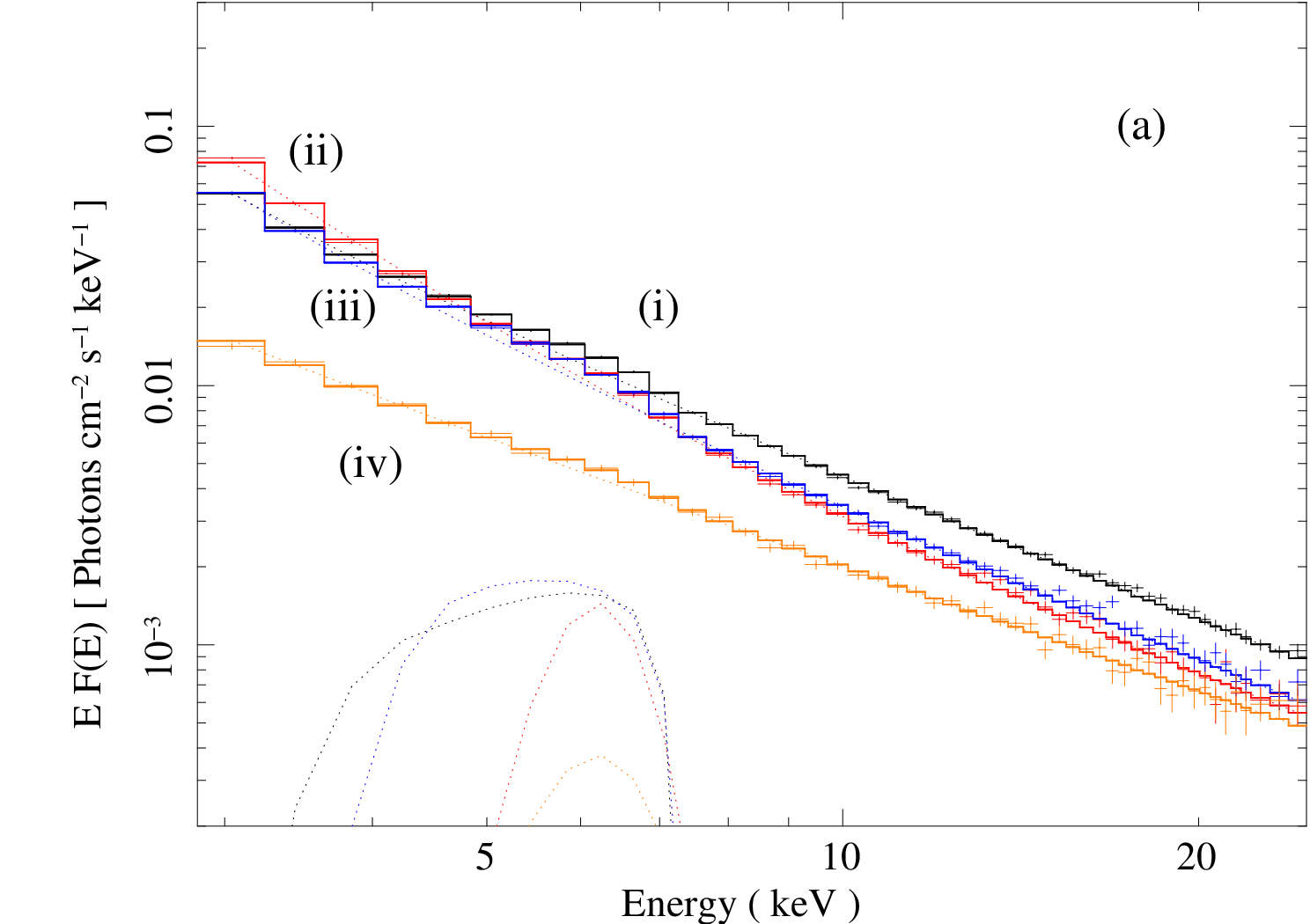}}\hskip 0.1cm
\vskip -0.5cm
\centerline{
\includegraphics[scale=0.6,angle=270,width=8.2truecm]{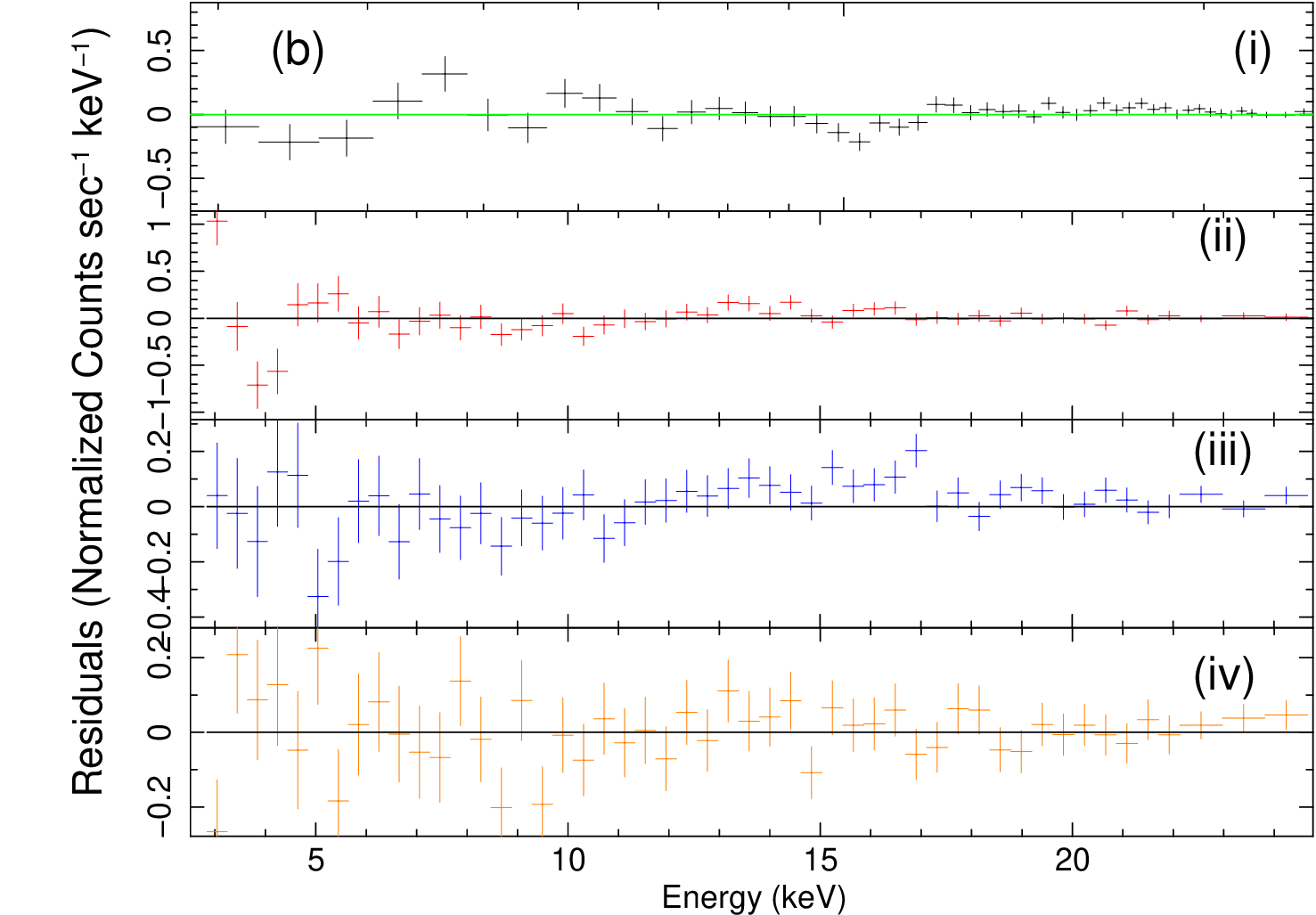}
}
\caption{In left panel (a), TCAF model fitted spectra for (i) HS (Ris.), (ii) HIMS (Ris.), (iii) HIMS (Dec.), 
and (iv) HS (Dec.) for observations Ids : 96371-03-03-01 (MJD=55808.34), 96438-01-01-04 (MJD=55818.85), 
96438-01-02-04 (MJD=55823.82), and 96438-01-06-02 (MJD=55850.87) respectively. In plots (i) \& (iii) we use 
an additional LAOR component and in (ii) \& (iv) a Gaussian Fe emission line are used to fit the spectra.
Observed data (points with error bars) and combined models (solid histogram) are shown in the plots. 
In the right panel (b), model fitted residual plots are shown. The parameters of the fits are listed 
in Appendix Table I.
}
\label{fig5}
\end{figure}

\begin{figure}
\vskip -0.5cm
\centerline{
\includegraphics[scale=0.6,angle=0,width=8.2truecm]{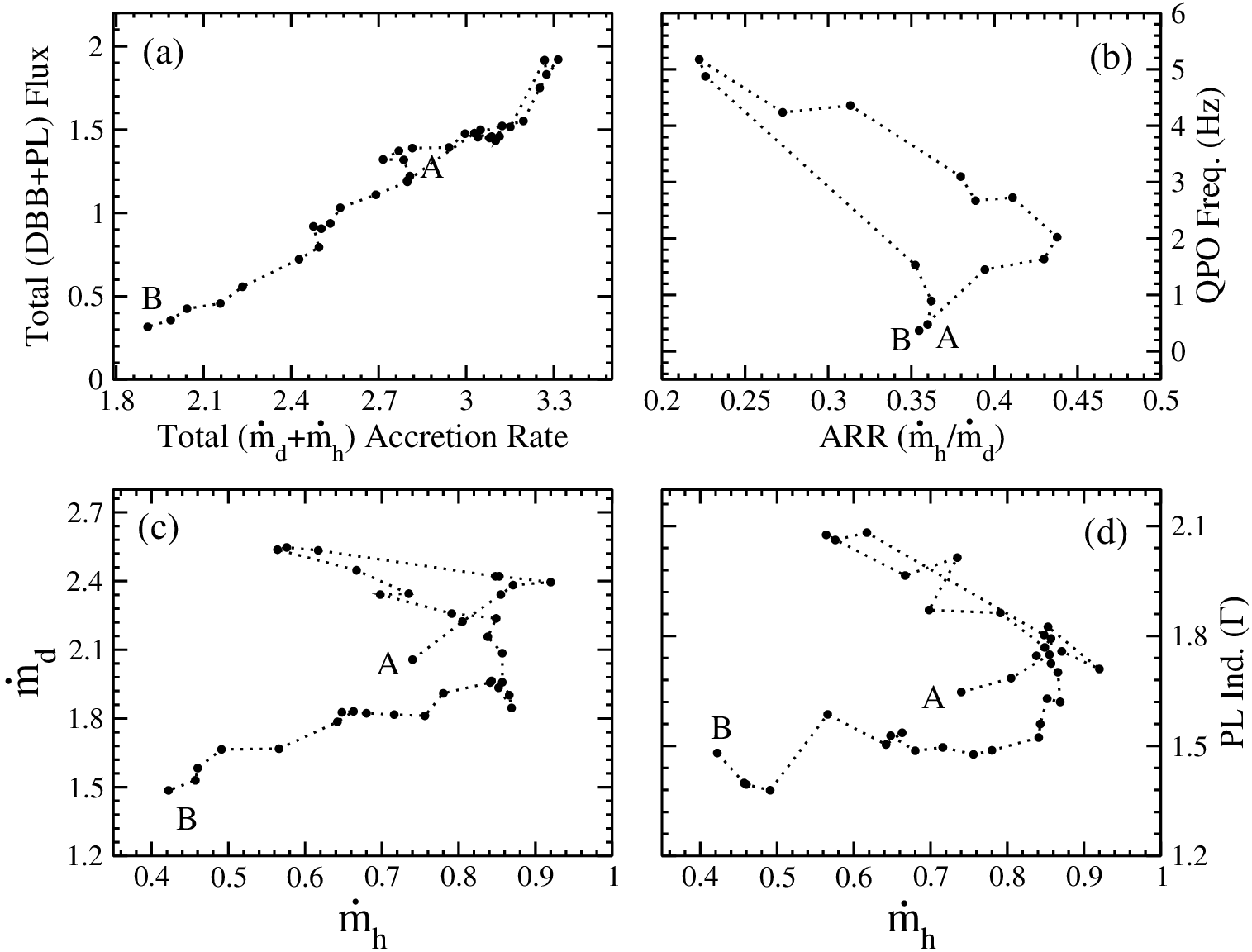}
}
\caption{Comparative plots between (a) total flow ($\dot{m_d}$ plus $\dot{m_h}$) rates vs. total (DBB plus PL) flux, 
(b) ARR vs. QPO frequencies, (c) $\dot{m_h}$ vs. $\dot{m_d}$ rates, and (d) $\dot{m_h}$ rates vs. PL photon indices. 
Note: here points A, and B are mark start and end of our observations.
}
\label{fig6}
\end{figure}

\clearpage

\begin{table}
\vskip -2.0cm
\addtolength{\tabcolsep}{-4.50pt}
\scriptsize
\centering
\centering{\large \bf Appendix I}
\vskip 0.2cm
\centerline {2.5-25 keV Combined DBB plus PL and TCAF Model Fitted Spectral Parameters with QPOs}
\vskip 0.2cm
\begin{tabular}{lcccccccccccccccc}
\hline
Obs. Id&MJD&$T_{in}$&$\Gamma$&$DBBf^\dagger$&$PLf^\dagger$&DBB Norm.&$\dot{m_d}$&$\dot{m_h}$&ARR&$X_s$&R&Norm.&$M_{BH}$&QPO$^{\dagger\dagger}$&$\chi^2/DOF$\\
    & & (keV) & & & &  &($\dot{M}$$_{Edd}$)&($\dot{M}$$_{Edd}$)&&($r_g$)& & & ($M_\odot$) & (Hz) &   \\
 (1)&  (2)  & (3)  & (4)& (5) & (6) & (7) &  (8) & (9) & (10) & (11) & (12) & (13)  & (14) & (15) & (16) \\
\hline
X-01-00$^*$   &55804.52&$1.061^{0.029}$&$1.647^{0.018}$&$0.115^{0.025}$&$1.078^{0.012}$&$11.27^{1.38}$&$2.057^{0.023}$&$0.740^{0.010}$&$0.360^{0.009}$&$  37.84^{3.820}$&$1.072^{0.017}$&$0.341^{0.002}$&$10.49 ^{0.22}$&$0.476^{0.027} $& 61.33/41\\
X-02-00$^*$   &55805.61&$1.066^{0.023}$&$1.685^{0.012}$&$0.164^{0.027}$&$1.315^{0.014}$&$15.63^{1.62}$&$2.223^{0.023}$&$0.805^{0.013}$&$0.362^{0.010}$&$  35.93^{2.230}$&$1.070^{0.004}$&$0.347^{0.002}$&$10.91 ^{0.34}$&$ ---          $& 62.41/41\\
X-03-00$^{**}$&55806.51&$0.963^{0.037}$&$1.749^{0.015}$&$0.188^{0.018}$&$1.364^{0.010}$&$31.25^{2.46}$&$2.341^{0.028}$&$0.853^{0.003}$&$0.364^{0.007}$&$  35.23^{1.511}$&$1.065^{0.005}$&$0.349^{0.027}$&$10.96 ^{0.34}$&$0.895^{0.050} $& 44.93/38\\
X-03-01$^{**}$&55808.33&$1.035^{0.035}$&$1.758^{0.018}$&$0.214^{0.015}$&$1.536^{0.011}$&$23.98^{1.94}$&$2.380^{0.012}$&$0.870^{0.011}$&$0.366^{0.007}$&$  39.00^{2.210}$&$1.057^{0.005}$&$0.741^{0.057}$&$9.99 ^{0.31}$&$ ---   	$& 28.58/38\\
 & & & & & & & & &  & & & & & & \\
X-03-02$^*$   &55810.29&$0.964^{0.021}$&$1.710^{0.015}$&$0.206^{0.018}$&$1.715^{0.011}$&$34.02^{1.98}$&$2.395^{0.004}$&$0.920^{0.005}$&$0.384^{0.003}$&$  34.19^{2.470}$&$1.053^{0.009}$&$0.345^{0.076}$&$10.87 ^{0.37}$&$ ---          $& 65.27/41\\
X-03-03$^{**}$&55812.57&$0.907^{0.018}$&$1.825^{0.019}$&$0.263^{0.012}$&$1.569^{0.012}$&$60.71^{1.66}$&$2.411^{0.004}$&$0.853^{0.005}$&$0.354^{0.003}$&$  39.01^{1.118}$&$1.057^{0.003}$&$0.542^{0.087}$&$11.01 ^{0.27}$&$1.531^{0.066} $& 53.50/38\\
Y-01-00$^*$   &55813.55&$0.971^{0.039}$&$1.803^{0.017}$&$0.179^{0.014}$&$1.738^{0.013}$&$280.24^{17.9}$&$2.421^{0.017}$&$0.848^{0.012}$&$0.350^{0.009}$&$  32.51^{2.510}$&$1.069^{0.014}$&$0.343^{0.017}$&$10.82 ^{0.38}$&$ ---  	$& 43.90/41\\
Y-01-04$^*$   &55818.84&$0.713^{0.021}$&$2.082^{0.028}$&$0.307^{0.015}$&$1.209^{0.015}$&$290.72^{17.0}$&$2.534^{0.021}$&$0.617^{0.003}$&$0.244^{0.004}$&$  47.00^{2.360}$&$1.053^{0.013}$&$2.073^{0.186}$&$10.63 ^{0.27}$&$ ---   	$& 52.75/41\\
Y-01-05$^*$   &55819.20&$0.753^{0.013}$&$2.062^{0.015}$&$0.302^{0.009}$&$1.220^{0.008}$&$207.75^{9.50}$&$2.547^{0.028}$&$0.576^{0.011}$&$0.226^{0.007}$&$  39.16^{2.450}$&$1.058^{0.015}$&$1.514^{0.187}$&$10.89 ^{0.23}$&$4.876^{0.085} $& 50.13/41\\
 & & & & & & & & &  & & & & & & \\
Y-02-03$^*$   &55820.40&$0.720^{0.014}$&$2.076^{0.018}$&$0.306^{0.009}$&$1.127^{0.009}$&$279.67^{9.07}$&$2.537^{0.014}$&$0.564^{0.010}$&$0.222^{0.006}$&$  39.95^{4.890}$&$1.057^{0.015}$&$1.682^{0.242}$&$10.89 ^{0.23}$&$5.175^{0.044} $& 72.70/41\\
Y-02-00$^*$   &55821.85&$0.890^{0.016}$&$1.964^{0.019}$&$0.294^{0.010}$&$1.164^{0.009}$&$63.73^{3.25}$&$2.447^{0.011}$&$0.667^{0.009}$&$0.273^{0.005}$&$  38.04^{2.470}$&$1.057^{0.016}$&$0.894^{0.015}$&$10.58 ^{0.33}$&$4.238^{0.215} $& 58.69/41\\
Y-02-01$^*$   &55822.83&$0.808^{0.027}$&$2.014^{0.027}$&$0.271^{0.015}$&$1.179^{0.014}$&$110.9^{1.80}$&$2.345^{0.026}$&$0.735^{0.008}$&$0.313^{0.007}$&$  37.82^{2.320}$&$1.055^{0.009}$&$0.799^{0.159}$&$10.98 ^{0.54}$&$4.359^{0.145} $& 73.73/41\\
Y-02-04$^{**}$&55823.81&$0.975^{0.027}$&$1.871^{0.025}$&$0.265^{0.011}$&$1.189^{0.013}$&$37.45^{2.31}$&$2.341^{0.024}$&$0.699^{0.002}$&$0.299^{0.003}$&$  41.02^{1.971}$&$1.058^{0.071}$&$0.541^{0.005}$&$10.91 ^{0.15}$&$ ---          $& 28.93/38\\
Y-02-05$^{**}$&55824.58&$0.895^{0.032}$&$1.863^{0.020}$&$0.184^{0.012}$&$1.315^{0.012}$&$45.70^{2.08}$&$2.259^{0.023}$&$0.791^{0.004}$&$0.350^{0.005}$&$  31.19^{1.214}$&$1.064^{0.014}$&$0.315^{0.031}$&$10.50 ^{0.13}$&$ ---  	$& 50.77/38\\
Y-02-02$^*$   &55825.94&$1.015^{0.029}$&$1.769^{0.019}$&$0.203^{0.012}$&$1.255^{0.011}$&$25.11^{1.90}$&$2.237^{0.046}$&$0.849^{0.000}$&$0.380^{0.008}$&$  37.80^{1.910}$&$1.060^{0.001}$&$0.504^{0.033}$&$10.86 ^{0.49}$&$3.099^{0.066} $& 67.03/41\\
Y-02-06$^*$   &55826.60&$1.113^{0.029}$&$1.746^{0.030}$&$0.208^{0.013}$&$1.267^{0.013}$&$14.13^{0.80}$&$2.157^{0.028}$&$0.838^{0.002}$&$0.389^{0.006}$&$  33.52^{2.350}$&$1.065^{0.010}$&$0.334^{0.108}$&$10.78 ^{0.11}$&$2.727^{0.034} $& 58.38/41\\
Y-03-04$^*$   &55827.33&$0.943^{0.025}$&$1.793^{0.026}$&$0.132^{0.015}$&$1.261^{0.014}$&$24.52^{2.67}$&$2.085^{0.028}$&$0.857^{0.010}$&$0.411^{0.011}$&$  34.59^{1.650}$&$1.060^{0.008}$&$0.339^{0.095}$&$11.01 ^{0.24}$&$2.672^{0.012} $& 48.70/41\\
Y-03-01$^*$   &55829.80&$1.130^{0.022}$&$1.725^{0.024}$&$0.120^{0.016}$&$1.270^{0.015}$&$8.32^{0.72}$&$1.958^{0.049}$&$0.857^{0.004}$&$0.438^{0.013}$&$  36.63^{1.380}$&$1.058^{0.003}$&$0.330^{0.086}$&$10.62 ^{0.15}$&$2.023^{0.073} $& 46.63/41\\
Y-03-05$^*$   &55830.89&$1.114^{0.025}$&$1.701^{0.030}$&$0.099^{0.021}$&$1.272^{0.018}$&$7.45^{0.41}$&$1.902^{0.036}$&$0.866^{0.005}$&$0.455^{0.011}$&$  39.09^{2.540}$&$1.051^{0.003}$&$0.331^{0.053}$&$11.00 ^{0.41}$&$ ---   	$& 51.52/41\\
 & & & & & & & & &  & & & & & & \\
Y-03-02$^*$   &55831.84&$1.128^{0.026}$&$1.620^{0.026}$&$0.096^{0.017}$&$1.225^{0.029}$&$6.60^{0.92}$&$1.846^{0.023}$&$0.869^{0.006}$&$0.471^{0.009}$&$  39.99^{1.780}$&$1.054^{0.006}$&$0.313^{0.050}$&$ 9.49 ^{0.69}$&$ ---   	$& 53.16/41\\
Y-03-03$^*$   &55832.81&$1.200^{0.025}$&$1.629^{0.020}$&$0.096^{0.013}$&$1.222^{0.011}$&$4.89^{0.31}$&$1.934^{0.029}$&$0.852^{0.008}$&$0.441^{0.011}$&$  41.39^{2.970}$&$1.063^{0.009}$&$0.321^{0.110}$&$10.70 ^{0.58}$&$ ---   	$& 34.85/41\\
Y-04-01$^*$   &55835.81&$1.195^{0.028}$&$1.560^{0.020}$&$0.076^{0.012}$&$1.145^{0.014}$&$3.95^{0.26}$&$1.964^{0.014}$&$0.843^{0.007}$&$0.429^{0.007}$&$  46.34^{3.280}$&$1.071^{0.004}$&$0.330^{0.031}$&$ 9.34 ^{0.88}$&$ ---   	$& 41.26/41\\
Y-04-02$^*$   &55836.78&$1.287^{0.021}$&$1.523^{0.025}$&$0.086^{0.014}$&$1.101^{0.013}$&$3.05^{0.20}$&$1.957^{0.015}$&$0.841^{0.009}$&$0.430^{0.008}$&$  46.40^{4.970}$&$1.073^{0.009}$&$0.298^{0.044}$&$ 9.49 ^{0.91}$&$1.636^{0.092} $& 34.95/41\\
Y-04-04$^*$   &55838.83&$1.347^{0.034}$&$1.488^{0.038}$&$0.085^{0.021}$&$1.023^{0.018}$&$2.42^{0.19}$&$1.910^{0.035}$&$0.780^{0.007}$&$0.408^{0.011}$&$  53.28^{1.750}$&$1.067^{0.021}$&$0.332^{0.025}$&$ 9.49 ^{0.78}$&$ ---          $& 36.53/41\\
Y-04-06$^*$   &55840.78&$1.270^{0.039}$&$1.477^{0.033}$&$0.080^{0.016}$&$0.951^{0.014}$&$3.04^{0.14}$&$1.812^{0.039}$&$0.756^{0.007}$&$0.417^{0.013}$&$  49.84^{2.120}$&$1.070^{0.012}$&$0.293^{0.007}$&$ 8.00 ^{0.77}$&$ ---  	$& 43.20/41\\
Y-05-00$^*$   &55842.79&$1.420^{0.038}$&$1.496^{0.037}$&$0.077^{0.019}$&$0.860^{0.019}$&$1.68^{0.13}$&$1.817^{0.033}$&$0.716^{0.011}$&$0.394^{0.013}$&$  50.91^{2.560}$&$1.070^{0.015}$&$0.280^{0.008}$&$ 8.79 ^{0.89}$&$1.449^{0.074} $& 58.38/41\\
Y-05-01$^*$   &55843.76&$1.468^{0.040}$&$1.487^{0.033}$&$0.079^{0.016}$&$0.827^{0.016}$&$1.46^{0.12}$&$1.823^{0.039}$&$0.680^{0.001}$&$0.373^{0.009}$&$  53.44^{2.840}$&$1.070^{0.012}$&$0.280^{0.057}$&$ 9.45 ^{0.52}$&$ ---   	$& 44.12/41\\
Y-05-04$^*$   &55844.86&$1.259^{0.034}$&$1.528^{0.039}$&$0.055^{0.018}$&$0.864^{0.015}$&$2.18^{0.18}$&$1.827^{0.022}$&$0.648^{0.009}$&$0.355^{0.009}$&$  57.40^{4.230}$&$1.071^{0.005}$&$0.323^{0.074}$&$ 9.45 ^{0.67}$&$0.368^{0.029} $& 40.58/41\\
Y-05-06$^*$   &55847.80&$1.237^{0.038}$&$1.536^{0.049}$&$0.032^{0.023}$&$0.762^{0.016}$&$1.39^{0.10}$&$1.831^{0.014}$&$0.663^{0.006}$&$0.362^{0.006}$&$  61.04^{2.470}$&$1.070^{0.009}$&$0.299^{0.042}$&$ 8.16 ^{0.87}$&$ ---   	$& 47.60/41\\
Y-06-02$^*$   &55850.87&$1.377^{0.037}$&$1.504^{0.046}$&$0.048^{0.027}$&$0.673^{0.021}$&$1.22^{0.08}$&$1.784^{0.017}$&$0.642^{0.002}$&$0.360^{0.004}$&$  60.38^{3.110}$&$1.069^{0.015}$&$0.269^{0.074}$&$8.26 ^{0.75}$&$ ---   	$& 35.71/41\\
Y-07-00       &55856.47&$1.194^{0.035}$&$1.586^{0.042}$&$0.016^{0.027}$&$0.541^{0.014}$&$0.77^{0.07}$&$1.668^{0.013}$&$0.566^{0.004}$&$0.339^{0.005}$&$  75.16^{1.740}$&$1.069^{0.008}$&$0.270^{0.027}$&$ 7.54 ^{0.41}$&$ ---   	$& 69.54/44\\
Y-08-00       &55862.87&$1.542^{0.030}$&$1.379^{0.034}$&$0.053^{0.023}$&$0.403^{0.020}$&$0.78^{0.06}$&$1.665^{0.015}$&$0.491^{0.002}$&$0.295^{0.004}$&$  96.45^{3.450}$&$1.076^{0.004}$&$0.251^{0.011}$&$ 7.51 ^{0.64}$&$ ---   	$& 52.25/44\\
Y-08-05       &55867.08&$1.474^{0.033}$&$1.395^{0.031}$&$0.042^{0.015}$&$0.384^{0.012}$&$0.76^{0.04}$&$1.583^{0.016}$&$0.460^{0.015}$&$0.291^{0.013}$&$ 135.48^{4.450}$&$1.084^{0.005}$&$0.251^{0.016}$&$ 7.72 ^{0.72}$&$  ---  	$& 51.72/44\\
Y-09-02       &55871.06&$1.649^{0.032}$&$1.399^{0.040}$&$0.038^{0.037}$&$0.318^{0.026}$&$0.40^{0.06}$&$1.530^{0.013}$&$0.457^{0.012}$&$0.299^{0.011}$&$ 162.37^{7.870}$&$1.084^{0.012}$&$0.250^{0.007}$&$ 7.76 ^{0.83}$&$ ---   	$& 65.29/44\\
Y-10-01       &55877.63&$1.612^{0.046}$&$1.481^{0.032}$&$0.020^{0.024}$&$0.296^{0.016}$&$0.34^{0.05}$&$1.486^{0.018}$&$0.422^{0.002}$&$0.284^{0.006}$&$ 199.24^{4.510}$&$1.089^{0.009}$&$0.258^{0.002}$&$ 7.57 ^{0.65}$&$ ---   	$& 64.36/44\\
                                                                                         
\hline

\end{tabular}
\noindent{
\leftline {X=96371-03 and Y=96438-01 are the prefixes of observation Ids. Blank lines mark transitions between different spectral states.}
\leftline {$^*$ Spectra are fitted with additional Gaussian lines of energy $\sim 6.5$~keV, and 
$^{**}$ spectra are fitted with an additional LAOR component with TCAF model.}
\leftline {$T_{in}$, and $\Gamma$ values indicate combined DBB and PL model fitted DBB temperatures in keV and PL photon indices respectively. }
\leftline {$^\dagger$ DBBf, PLf represent combined DBB and PL model fitted fluxes in 2.5-25~keV band for DBB and PL model components respectively in units of $10^{-9}~ergs~cm^{-2}~s^{-1}$. }
\leftline {$\dot{m_h}$ (in Eddington rate unit), $\dot{m_d}$ (in Eddington rate unit), $X_s$ (in units of Schwarzschild radius), and $R$ are TCAF fitted Keplerian disk, sub-Keplerian halo,} 
\leftline {shock location and compression ratio values respectively. Norm. and $M_{BH}$ are TCAF fitted normalization and BH mass values respectively.} 
\leftline {$^{\dagger\dagger}$ Frequencies of the dominating QPO in Hz are mentioned. DOF means degrees of freedom of the spectral model fits.}
\leftline {The values of $\chi^2$, and DOF of TCAF model fitted spectra are mentioned in Col. 16. The ratio is the $\chi^2_{red.}$.}
\leftline {Note: average values of 90\% confidence $\pm$ values obtained using `err' task in XSPEC, are placed as superscripts of fitted parameter values.}
}
\end{table}

\begin{table}
\vskip -2.0cm
\addtolength{\tabcolsep}{-4.50pt}
\scriptsize
\centering
\centering{\large \bf Appendix II}
\vskip 0.2cm
\centerline {Combined LAOR plus TCAF Spectral Fitted LAOR Parameters}
\vskip 0.2cm
\begin{tabular}{lccccccc}
\hline
obs. Id&MJD&Line Energy&Index&$R_{in}$&$R_{out}$&Inclination&LAOR Norm.\\
& &  (keV)  & &($r_g)$ &($r_g$)& (Degree)  &    \\
 (1)&  (2)  & (3)  & (4)& (5) & (6) & (7) & (8) \\
\hline

\hline
X-03-00&55806.51&$7.14^{0.08}$&$3.49^{0.02}$&$3.02^{0.61}$&$358.154^{9.97}$&$4.00^{0.03}$&$0.0042^{0.0001}$\\
X-03-01&55808.33&$7.11^{0.12}$&$3.50^{0.02}$&$3.04^{0.57}$&$330.13^{12.79}$&$4.12^{0.02}$&$0.0049^{0.0001}$\\
X-03-03&55812.57&$7.15^{0.10}$&$3.47^{0.02}$&$3.59^{0.48}$&$378.62^{14.21}$&$4.10^{0.04}$&$0.0047^{0.0002}$\\
Y-02-04&55823.81&$7.17^{0.14}$&$3.62^{0.02}$&$3.33^{0.51}$&$351.10^{11.71}$&$4.07^{0.03}$&$0.0046^{0.0004}$\\
Y-02-05&55824.58&$7.04^{0.18}$&$3.61^{0.14}$&$3.24^{0.42}$&$297.32^{10.33}$&$4.09^{0.07}$&$0.0054^{0.0002}$\\

\hline
\end{tabular}
\noindent{
\leftline {X=96371-03 and Y=96438-01 are the prefixes of observation Ids.} 
\leftline {Note: average values of 90\% confidence $\pm$ values obtained using `err' task in XSPEC, are placed as superscripts of fitted parameter values.}
}
\end{table}
\end{document}